\documentclass[aps,pre,twocolumn,groupedaddress,showpacs]{revtex4}
\newif\ifpdf            
\ifx\pdfoutput\undefined \pdffalse \else \pdftrue \fi
\ifpdf 
\usepackage[pdftex]{graphicx}
\pdfcompresslevel9 \DeclareGraphicsExtensions{.jpg,.pdf,.png,.mps}
\else 
\usepackage[dvips]{graphicx}
\DeclareGraphicsExtensions{.eps,.ps} \fi
\usepackage{bm}

\begin{document}

\title{Evolution of photonic structure on deformation of cholesteric elastomers}

\author{P.~Cicuta}
\author{A.R.~Tajbakhsh}
\author{E.M.~Terentjev}

\affiliation{Cavendish Laboratory, University of Cambridge,
Madingley Road, Cambridge, CB3 0HE, U.K. }

\date{\today}

\begin{abstract}
We subject a monodomain cholesteric liquid crystal elastomer to
uniaxial strain perpendicular to its helical axis and study the
response of its texture to deformation. A combination of
mechanical, optical and X-ray scattering measurements confirms the
prediction for the director rotation, coarsening and then
unwinding the cholesteric helix. The study of optical absorption
of circularly polarised light quantifies the complex dependence of
the photonic bandgap structure on strain and directly relates to
the microscopic deformation of elastomer. Agreement is found with
the recently proposed theoretical prediction of the photonic
structure of cholesteric elastomers.
\end{abstract}

\pacs{61.41.+e, 42.70.Qs, 83.80.Xz}

\maketitle

\section{Introduction}
Liquid crystal elastomers (LCE) are materials displaying both a
broken orientational symmetry and a sparsely crosslinked rubbery
network of flexible polymers. They have novel mechanical behaviour
arising from the coupling of the orientational degrees of freedom
of the liquid crystalline order and those of the rubber-elastic
matrix. In particular, soft deformations and unusual thermal and
photo-actuation have been discovered and studied in the aligned
monodomain nematic phase of rubbers \cite{WT96,ratna,uv}.

Cholesteric liquid crystals have attracted much interest over the
years due to their exciting optical properties \cite{dGP93}.
Cholesterics have also found use in very diverse products ranging
from thermometers to fabric dyes. There has been an effort for
some time to produce and study well-aligned cholesteric liquid
crystal polymers and their crosslinked networks -- permanently
stabilised Cholesteric Liquid Crystal Elastomers (CLCE). Much
success has been achieved in developing cholesteric polymers
\cite{chol1,chol2,chol3}, aligning them by electric and magnetic
fields and by surface interactions with substrates. However, the
permanent chemical crosslinking of cholesteric textures and
obtaining a freely-standing, mechanically stable rubbery films of
such polymers has proven much more difficult. Early experiments
achieved cholesteric order through the mechanical deformation of
an elastomer doped with a low mass chiral liquid crystal
\cite{MF93}. A second approach has been to form a perfect
cholesteric texture between parallel cell surfaces and then
forming a crosslinked network by UV-polymerisation
\cite{hasson,CCM97}. While these approaches led to materials with
interesting properties and enabled for example the study of
piezoelectric effects \cite{MF93, CCM97}, they are not ideal. In
the first method, the alignment is only induced mechanically and
it is not a permanent property of the material, while in the
second method only very thin films elastomers can be synthesized
and they cannot be used as freely standing samples. Recently a new
method of synthesis was devised which solves both problems. Free
standing strips of single-crystal, well aligned cholesteric rubber
have been produced by Kim and Finkelmann \cite{LCSE} and a uniform
orientation of helical director texture has been confirmed by
selective reflection measurements. The key idea of this new method
is the use of anisotropic (uniaxial) deswelling induced in a
laterally constrained film of a weakly crosslinked cholesteric
polymer; the resulting effective biaxial planar extension leads to
macroscopic orientation of the director in the plane and,
therefore, the helical axis perpendicular to the sample plane.
This orientation is then locked by a second-stage cross-linking of
the network.

\begin{figure}
\resizebox{0.42\textwidth}{!}{\includegraphics{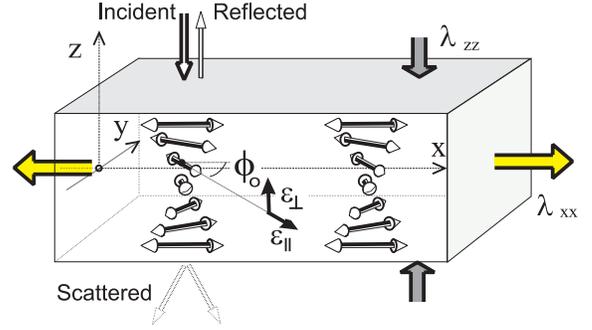}}
\caption{Diagram of the CLCE geometry showing the cholesteric
helical texture characterised by the uniform director rotation
about the pitch axis $z$. In our experiments the uniaxial strain
$\lambda=\lambda_{xx}$ is imposed along $x$; the associated
contraction of sample thickness is $\lambda_{zz}$. The incident
light and X-ray beams are sent parallel to the pitch axis,
resulting in the selective reflection and transmission for light
and scattering for X-rays.} \label{diagram}
\end{figure}

In a cholesteric liquid crystal the periodicity of helical
structure (characterised by the pitch $p$ and the wavenumber
$q_0=\pi/p$), combined with the local birefringence of the liquid
crystal ($\Delta n$), forbids propagation of circularly polarised
light with the corresponding handedness and the wavelength
$\Lambda_{gap}= p\,\Delta n$, determining a photonic bandgap. This
has been first observed in an elastomer in \cite{hasson,LCSE} as
the selective reflection of light. The remarkable property of
cholesteric elastomers is that $\Lambda_{gap}$ is tunable,
responding to mechanical deformation, as it has been theoretically
predicted in \cite{WTMM00} and experimentally observed in
\cite{laser}. These materials have many potential applications as
novel optical filters and mirrors. It has been demonstrated how a
dye-doped CLCE can form the basis of a mirrorless laser whose
wavelength can be controlled through mechanical deformation
\cite{laser}.

Theoretical work, describing CLCE and their response to mechanical
deformation, predicted a number of new effects, including a very
rich and complex evolution of photonic band structure
\cite{WTM01,BW01}. Briefly, when a rubber with cholesteric liquid
crystalline  microstructure is stretched uniaxially in the
$x$-direction, perpendicular to the helix axis $z$, as in
Fig.~\ref{diagram}, two main phenomena are expected to occur.
First, there is a contraction in the $z$ direction, due to the
rubber incompressibility, whose detail depends on the complex
transverse anisotropy of the cholesteric helix. Second, the
uniform helical rotation of the optical axis (the local nematic
director) is disrupted and the helix coarsens, thus producing
complex non-linear corrections to the resulting optical response.

In this paper we study experimentally the effect of uniaxial
strain on a CLCE. In marked contrast to the previously studied
bi-axial strain \cite{laser}, a uniaxial deformation does not
simply  reduce the helical pitch but produces a complex photonic
structure that modifies with deformation. We present optical
absorption and X-ray diffraction measurements on a CLCE sample as
a function of strain and and discuss the complex behavior that is
induced even by small uniaxial deformations.

\section{Methods}
\subsection{Materials and preparation}
\begin{figure}
\resizebox{0.42\textwidth}{!}{\includegraphics{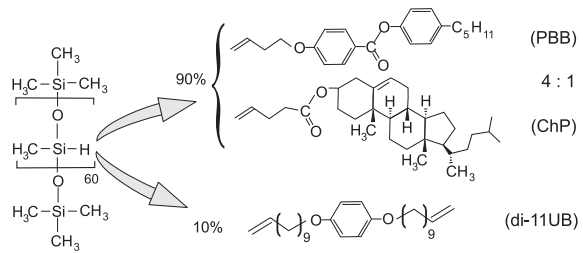}}
\caption{Chemical structures of the compounds forming the
cholesteric elastomer under investigation. A siloxane backbone
chain reacting with 90\,mol\% mesogenic side groups and 10\,mol\%
of the flexible di-functional crosslinking groups (di-11UB). The
rod-like mesogenic groups are divided in the proportion 4:1
between the nematic 4-pentylphenyl-4'-(4-buteneoxy)benzoate (PBB)
and the derivative of chiral cholesterol pentenoate (ChP). }
\label{chemistry}
\end{figure}

A new CLCE was synthesized following the general method introduced
by Kim and Finkelmann \cite{LCSE}. Siloxane backbone chains were
reacted under centrifugation at 7000rpm with 90\,mol\% mesogenic
side groups (the nematic 4-pentylphenyl-4'-(4-buteneoxy)benzoate,
labelled PBB, and Cholesterol Pentenoate, ChP, in proportion 4:1)
and 10\,mol\% of 1,4\,di(11-undeceneoxy)benzene, di-11UB,
crosslinker groups for 45 minutes at 75$^\circ$C to form a
partially crosslinked gel. For the further 4 hours the reaction
proceeded under centrifugation at 60$^\circ$C, during which time
the solvent was allowed to evaporate, leading to an anisotropic
deswelling of the gel and completion of crosslinking. All of the
volume change in this setup occurs by reducing the thickness of
the gel, while keeping the lateral dimensions fixed (due to
centrifugation): this introduces a very strong effective biaxial
extension in the plane. At this second-stage temperature of
60$^\circ$C the dried polymer is in the cholesteric phase and its
director is forced to remain in the plane of stretching -- this
results in the uniform cholesteric texture. The structure of the
material produced is shown in Fig.~\ref{chemistry}. This kind of
synthesis produces a very homogeneous strip of elastomer whose
size is of the order of 20cm$\times$1cm$\times$200$\mu$m.
Differential scanning calorimetry (DSC) measurements (Perkin-Elmer
Pyris 7 DSC) were used to characterise the resulting elastomer.
The glass transition was unambiguously determined at $T_g \approx
-10^\circ$C and the clearing point, the isotropic-cholesteric
transition occurs at $T_{\rm c}\approx 90^\circ$C. No additional
thermal transitions were found between these two critical
temperatures. All experiments were performed at room temperature,
sufficiently far from both transitions.

\subsection{Mechanical measurements}
The stress-strain measurements were performed on a custom built
device consisting of a temperature compensated stress gauge and
controller(UF1 and AD20, from Pioden Controls Ltd) in a
thermostatically controlled chamber (Cal3200 from Cal Controls
Ltd). The long elastomer strips were mounted with clamps and
extended with a micrometer. The resulting force (as well as the
current temperature) values were acquired by connection to a
Keithley multimeter (2000 series) and stored on a PC over an IEEE
interface. Data obtained in arbitrary units were converted to
nominal stress in units of mN/mm${}^2$=$10^3$Pa by calibration
with weights.

\subsection{X-Ray Scattering}
Wide angle X-ray scattering measurements were performed at room
temperature with mono-chromatic CuK$_\alpha$-radiation of 0.154nm,
using a two-dimensional photographic plate. Images were then
digitalized and analyzed to extract the azimuthal distribution of
intensity indicating the degree of uniaxial order. As with the
optical measurements, the beam was incident in the cholesteric
pitch direction $z$, and measurements were performed as a function
of applied uniaxial strain $\lambda_{xx}$.

\subsection{Visible Light Absorbance}
\begin{figure}[h]
\resizebox{0.42\textwidth}{!}{\includegraphics{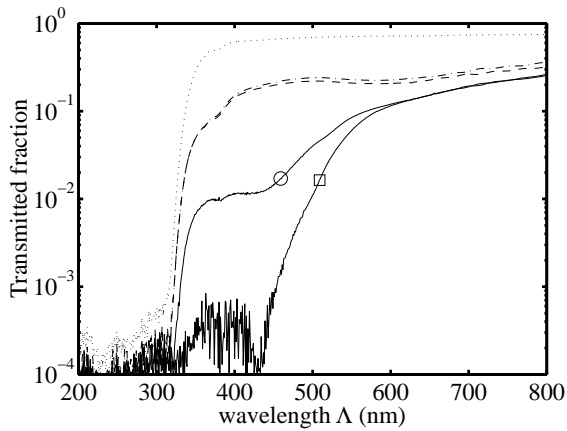}}
\caption{The absorption due to the Fresnel rhomb (dots) and the
baseline absorption due to the complete circular polariser
assembly for L* (dashed line) and R* (dash-dot line)
polarisations. The fraction of transmitted light ($I/I_0$,
logarithmic scale) for the equilibrium unstretched monodomain CLCE
is shown by solid lines, labelled by the circle for L* and by the
square for R* circular polarisations.} \label{figureabss2}
\end{figure}

Absorption measurements in the direction of the cholesteric pitch
(Fig.~\ref{diagram}) were made at room temperature with an HP-8453
UV/Visible spectrophotometer, modified so that incident light
reached the sample after being transmitted through a linear
polarizer and a Fresnel rhomb. Since the Fresnel rhomb is an
achromatic quarter-wave retarder, this arrangement produces
circularly polarized light throughout the visible spectrum
\cite{optics} with minimal losses of intensity. The handedness of
circular polarization is determined by the orientation of the
linear polarizer. An instrumental cut-off, due to both the
polarizer and the glass prism, is apparent at $\sim$320nm
(Fig.~\ref{figureabss2}). We have measured the transmitted light,
and hence will present the data as the ratio $I/I_0$ of
transmitted to incident light. It is a property of the cholesteric
materials under study that the fraction of light that is not
transmitted is not absorbed by the sample, but mostly reflected
backwards. We comment that the direct measurement of the reflected
beam would reduce any effects of scattering from the sample, but
would not extend the accessible range to the UV where the
elastomer is not transparent.

\section{Results}

The first, most intuitively obvious effect occurring on uniaxial
stretching of a monodomain cholesteric elastomer across its
helical axis is the gradual unwinding of the director helix. As
the theoretical model \cite{WTMM00} predicted for liquid
crystalline polymers with prolate anisotropy of the backbone
(chain preferentially extended along the local director), on
application of uniaxial strain the director tends to align along
the stress axis. This is resisted by the anchoring of the local
director in the bulk of the material to the initial helical
texture frozen-in by the network crosslinks. We first examine the
mechanical response of such a system and analyse the director
distribution from the X-ray scattering data. Having qualitatively
confirmed the theoretical prediction of the helix coarsening and
eventual unwinding via a continuous set of highly non-uniform
director textures, we then study how these affect the photonic
band structure of the material.

\subsection{Stress-strain response}
The equilibrium stress-strain curve of cholesteric rubber
stretched perpendicular to its helix axis could be expected to
possess anomalies, since the complex pattern of internal director
rotation should follow the deformation; this often is a
requirement for soft-elastic response. However, the detailed
theoretical calculation \cite{WTM01} has shown that in this
geometry (cf. Fig.~\ref{diagram}) the rotations are not soft. In
fact, only a small anomaly is predicted (within this ideal
molecular model) at the critical point of transition where the
coarsened winding texture transforms into the non-helical
coarsened wagging one. The estimate of critical strain for this
discontinuous transition, $\lambda_{\rm c} = r^{2/7}$, is related
to the effective parameter of backbone anisotropy $r$. This is the
single parameter of the ideal theory of nematic (and cholesteric)
elastomers; $r=\ell_\|/\ell_\bot$ is interpreted as the average
ratio of backbone chain step lengths along and perpendicular to
the local director. $r$ could be directly measured by examining
the uniaxial thermal expansion of nematic rubber. Recently much
attention has been paid to evaluating this parameter
\cite{greve,static} and directly relating it to the local value of
underlying nematic order parameter $Q(T)$. To a good approximation
a relationship $r \approx (1+ \alpha \, Q)^3$ holds with a
coefficient $\alpha$ taking a value $\alpha \sim 0.9-1$ for the
side-chain siloxane liquid crystalline elastomer crosslinked by
the flexible linkages, as shown in Fig.~\ref{chemistry}.

\begin{figure}
  \resizebox{0.42\textwidth}{!}{\includegraphics{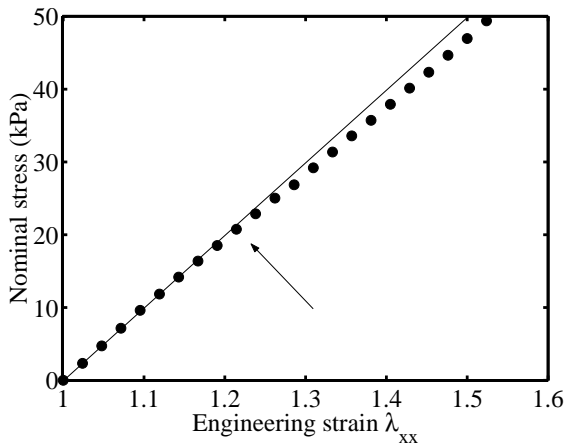}}
\caption{Stress-strain response of monodomain CLCE stretched
uniaxially in the direction perpendicular to the helix. The value
of linear rubber modulus at small deformations is $\mu \approx
89~\hbox{kPa}$. The arrow indicates the breaking point of the
initial linear regime; its strain $\lambda^* \sim 1.25$.}
\label{stress}
\end{figure}

Figure~\ref{stress} shows the experimentally measured
stress-strain curve for our CLCE. The mechanical response is
essentially linear for strains up to 20-25\%. In this regime the
linear rubber modulus is $\mu \approx 89~\hbox{kPa}$, consistent
with other measurements performed on nematic materials with
similar composition and crosslinking density. There is a break in
the linear stress-strain relation, emphasized by the arrow in the
plot. We could attribute this to the theoretically predicted
anomaly \cite{WTM01}, when at a critical strain $\lambda_{\rm c}$
the stress drops below the linear regime. If so, then the value of
local backbone anisotropy $r = \lambda_{\rm c}^{7/2} \approx 2.2$,
which is not far from the typical values measured in similar
nematic rubbers ($r$ between $2.5$ and $3$ are reported for such
materials). The little discrepancy in the values of $r$ is to be
expected: the usual monodomain nematic rubbers are prepared by
two-step crosslinking with the uniaxial strain of up to 300\%,
while the cholesteric monodomain was prepared under the effective
biaxial strain of much lower magnitude -- hence, perhaps, the
slightly lower value of effectively frozen chain anisotropy.

\subsection{Helix coarsening and unwinding}

\begin{figure}
\resizebox{0.42\textwidth}{!}{\includegraphics{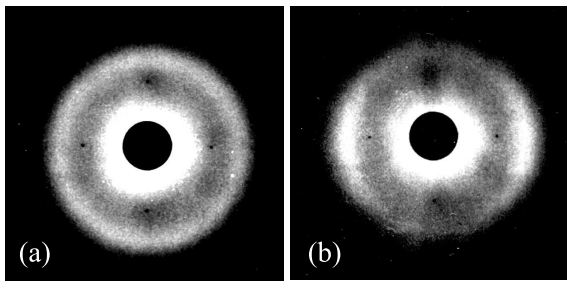} }
 \caption{Two typical X-ray scattering patterns obtained before
deformation: (a)~$\lambda_{xx}=1$, and at a relatively high
imposed strain: (b)~$\lambda_{xx}=1.4$ (40\% extension). In the
first case there is no azimuthal bias from the perfect helical
texture viewed along its axis, while the stretched material shows
a high degree of average alignment along the stress axis. }
\label{Xphotos}
\end{figure}

\begin{figure}
\resizebox{0.42\textwidth}{!}{\includegraphics{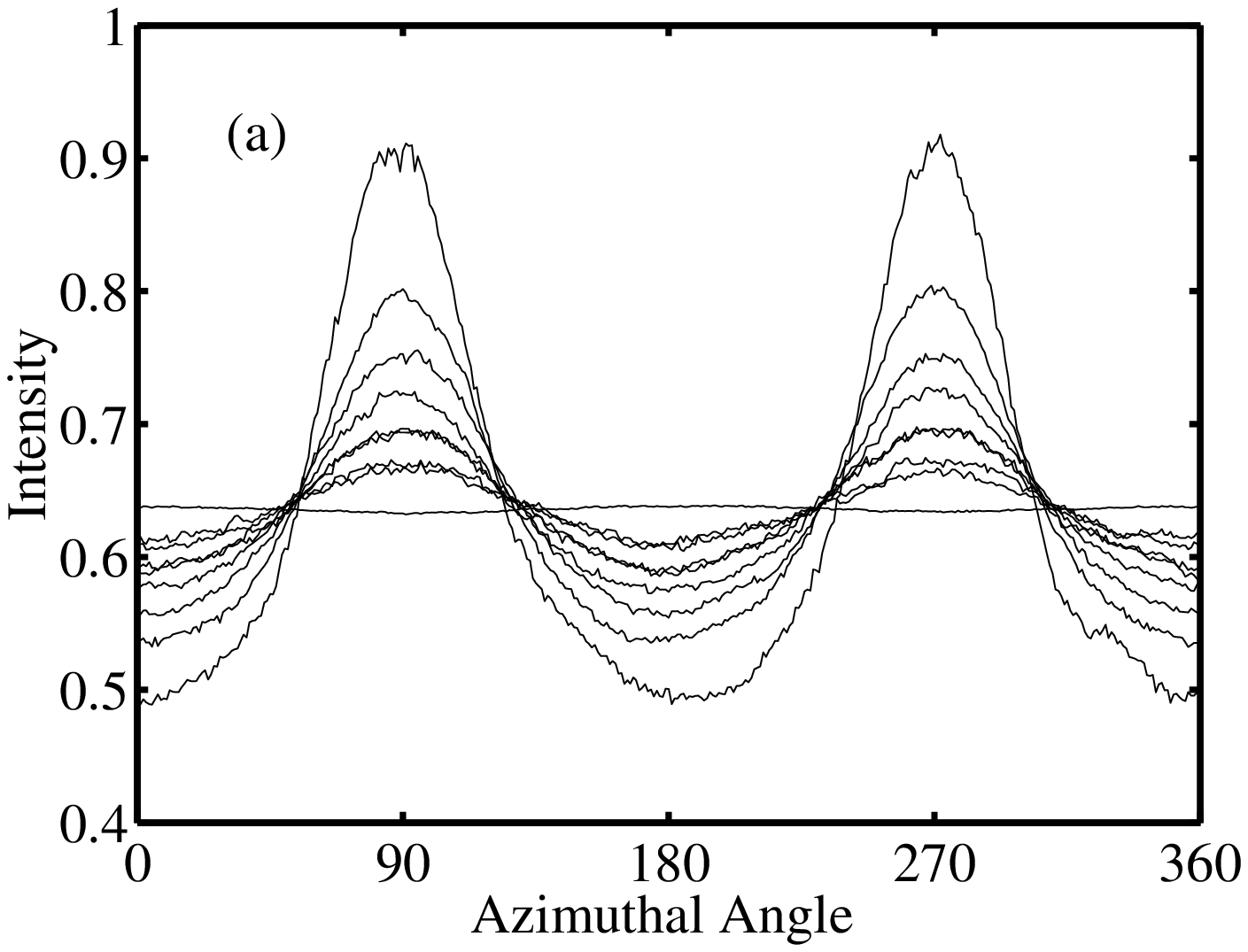} }\\
\resizebox{0.42\textwidth}{!}{\includegraphics{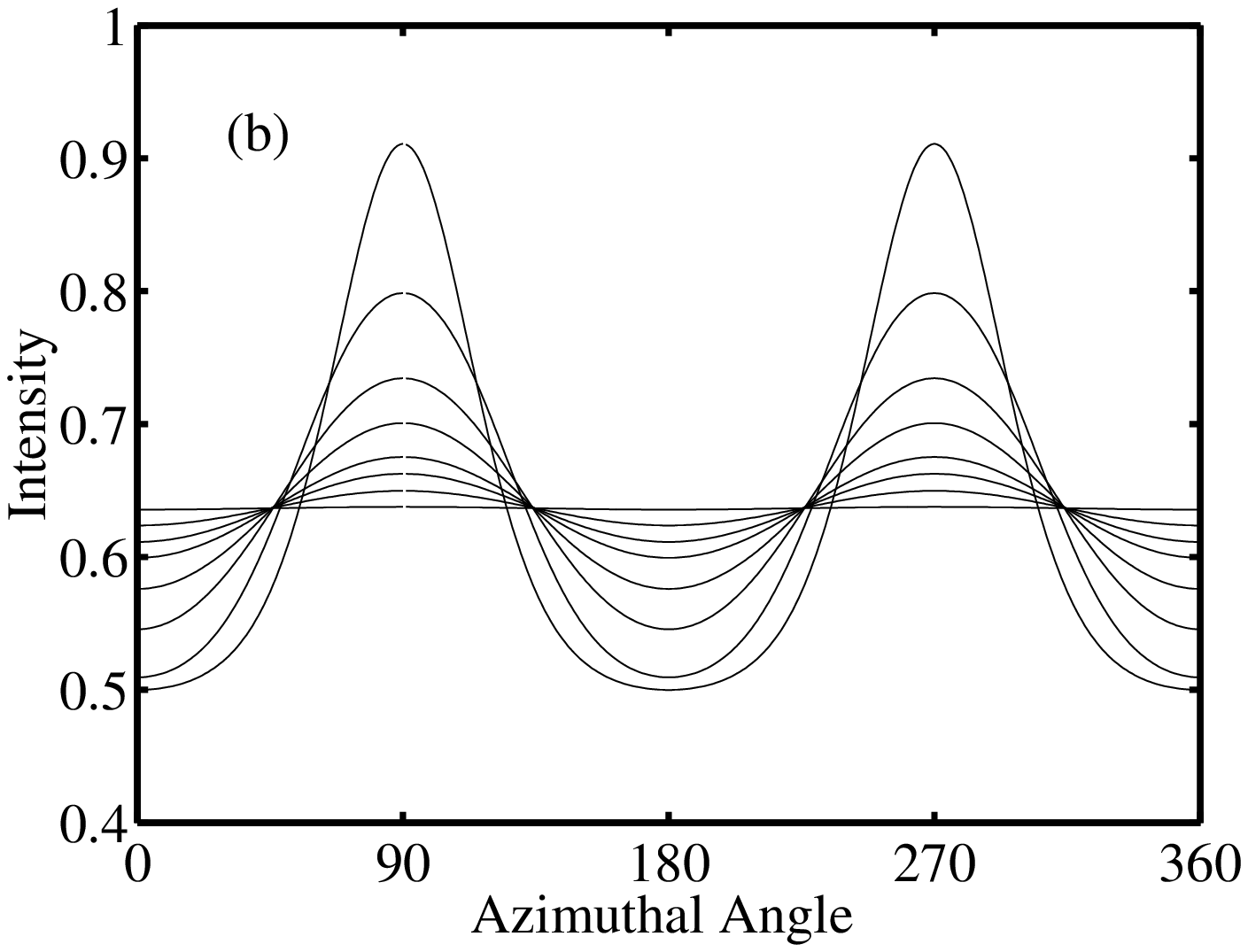} }
\caption{(a) Azimuthal intensity distribution around the
wide-angle diffraction ring for $\lambda_{xx}=1.00, 1.04,   1.08,
1.12, 1.16, 1.24, 1.32$ and $1.40$. At $\lambda_{xx}=1.00$, the
profile is flat to within our experimental error. The scattering
profile showing the highest dependance of  intensity on azimuthal
angle is obtained from a nematic liquid crystal elastomer.  (b)
Theoretical prediction from Eq.~\ref{phi} for the same strain
values as (a).} \label{xrayprofile}
\end{figure}

Fig.~\ref{Xphotos} shows X-ray scattering patterns for the
equilibrium (unstretched) cholesteric rubber (a) and the same
rubber stretched by $\lambda_{xx}=1.4$ (b). In this scattering
geometry, the wide-angle diffraction is due to the characteristic
dimension given by the thickness of the aligned and densely packed
mesogenic rods \cite{Deu91}, averaged over the local director
rotating along the $z$-axis. The scattered intensity around this
ring is initially uniform, see Fig.~\ref{Xphotos}(a), indicating
that the nematic director is uniformly distributed along $z$ (the
angle $\phi(z)=q_0z$, cf. Fig.~\ref{diagram}). As the sample is
stretched, azimuthal lobes develop, see Fig.~\ref{Xphotos}(b),
indicating that there is now a preferred orientation of the
nematic director, along the direction of elongation. There is no
detectable threshold for lobes  to develop, as can be seen from
the azimuthal intensity scans shown in Fig.~\ref{xrayprofile}(a).

We can, in fact, model these azimuthal intensity profiles by
assuming that the cholesteric structure is locally exactly the
same as that of a chemically similar nematic and then performing
the average along the helical pitch. To do so, we need the X-ray
scattering data for a corresponding monodomain (aligned) nematic
elastomer -- which is presented for comparison as the maximal
curve in Fig.~\ref{xrayprofile}(a); this gives the reasonable
value of nematic order parameter $Q \approx 0.5$, calculated by
standard methods \cite{Deu91}. One then assumes a stack of such
nematic layers along the beam path, continuously rotating by the
angle $\phi(z)$. In an ideal cholesteric $\phi=q_0z$ and the
predicted azimuthal distribution is uniform: all director
orientations in the plane are equally represented in the
scattering.

As soon as an external strain $\lambda=\lambda_{xx}$ is applied,
the cholesteric texture is modified, with its angle expressed by
the formula derived in \cite{WTMM00}:
\begin{equation}
\tan 2\phi =\frac{2\lambda^{1/4} (r-1) \sin \, 2 \tilde{q}z}
{(r-1)(\lambda^2+\lambda^{-3/2})\cos \,2 \tilde{q}z
+(r+1)(\lambda^2- \lambda^{-3/2}) } \label{phi}
\end{equation}
This expression is written in a slightly different format than in
\cite{WTMM00}, using a shorthand $\lambda$ for the imposed
extension $\lambda_{xx}$ and substituting the calculated value for
the transverse sample contraction due to incompressibility,
$\lambda_{zz}= \lambda^{-1/4}$. Also, the corresponding affine
contraction of the cholesteric pitch is taken into account by
taking the wavenumber $\tilde{q}=\lambda^{1/4}q_0$. Eq.
(\ref{phi}) expresses the strain-induced bias of director
orientations along the $x$-direction: the coarsening of the helix
occurs (interrupted by the critical jump at $\lambda_{\rm c}$
which is the point of zero denominator of $\tan 2\phi$). Taking
the value of $r$ as estimated from the stress-strain data in
Fig.~\ref{stress} and the values of imposed strain $\lambda$ as
labelled in Fig.~\ref{xrayprofile}(a), we can analytically predict
how the azimuthal intensity distribution would evolve -- see
Fig.~\ref{xrayprofile}(b). The agreement is very satisfactory,
suggesting that the theoretical description of helix coarsening in
CLCE is reasonable.

\subsection{Effects of strain on the photonic structure}
It is evident from a comparison of the unstretched sample  spectra
in Fig.~\ref{figureabss2}  how  the sample transmits very
differently light of  opposite polarizations. In the region of
wavelengths around $\Lambda=$400nm a difference in transmission
greater than 2\,O.D. is measured \footnote{The optical density
(O.D.) is a unit measuring the effective attenuation of
transmitted light, on logarithmic scale, so that a difference of
2\,O.D. between R* and L*  transmitted light intensity means a
ratio of 100.}.

\begin{figure}
\resizebox{0.42\textwidth}{!}{\includegraphics{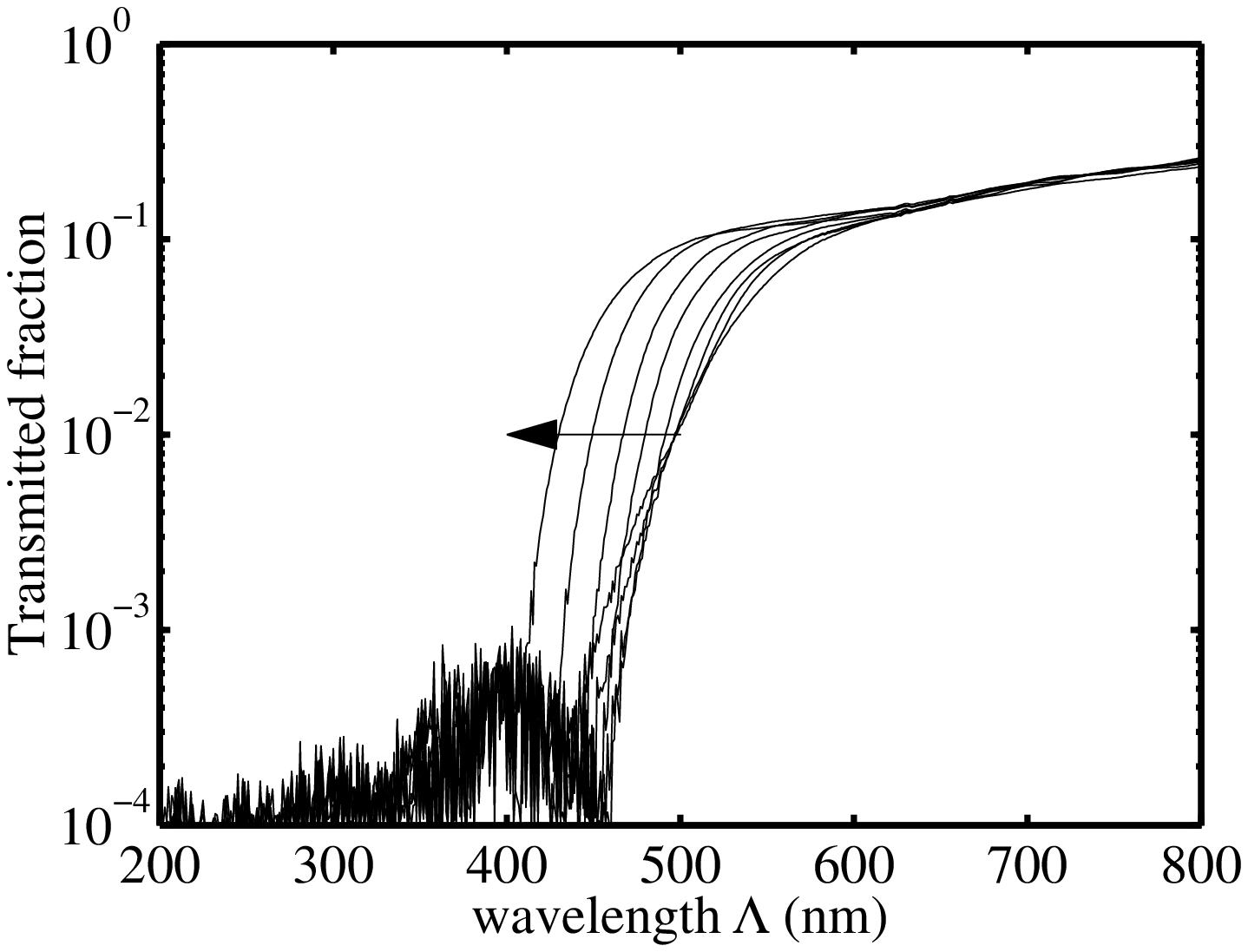}}\\
\resizebox{0.42\textwidth}{!}{\includegraphics{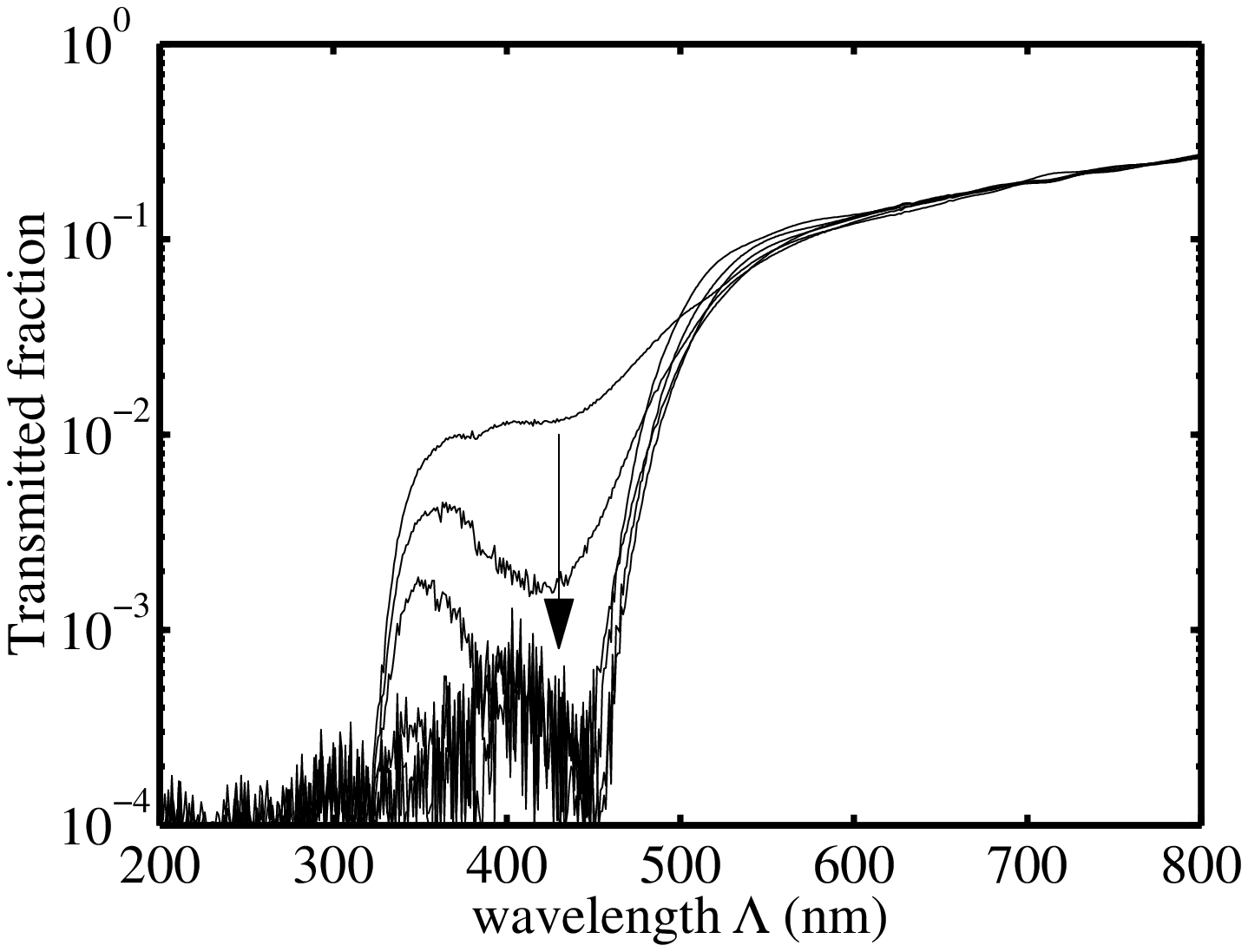}}
\caption{(a) Helicity R*. On stretching, the main R* reflection
remains but the line edge sharpens and moves to shorter
wavelengths. The spectra are presented at consecutive strains
$\lambda_{xx} = 1, 1.02,   1.04,   1.08,    1.16,   1.24, 1.40,
1.64$, the arrow indicating increasing strain. \ (b) Helicity L*.
A dip in the transmitted intensity (the new gap) develops under
strain, until the sample becomes opaque at that wavelength. The
spectra are presented at consecutive strains$\lambda_{xx} = 1,
1.02, 1.04,  1.08,  1.12, 1.16$, the arrow indicating increasing
strain.} \label{figureabsb}
\end{figure}

Figs.~\ref{figureabsb}(a,b) show the evolution of absorption
spectra of right and left hand circularly polarized light,
respectively, as the sample is uniaxially strained. The data in
Fig.~\ref{figureabsb}(a) clearly shows that with no applied strain
R* polarized light is reflected  at wavelenghts below $\sim$500nm.
This gap  wavelength shifts downwards as the sample is stretched,
as shown in Fig.~\ref{strain}.
The transmitted intensity  spectra of L* polarized light
Fig.~\ref{figureabsb}(b) reveal a dramatic development of a new
reflection gap centered around $\sim$430nm as soon as the sample
is stretched. With further increasing uniaxial strain, at $\lambda
> 1.04$ the L* reflection gap appears to have fine structure, the
transmitted intensity shows two separate minima respectively at
$\sim$390nm and $\sim$440nm.  We remark that as no R* polarized
light is transmitted here, this complementary nearly-total
reflection of L* polarized light means that for
$\lambda_{xx}>1.04$ the material is not transmitting any light at
the wavelengths corresponding to the L* reflection peak. We have
measured spectra of transmitted non-polarized light that confirm
this. Above a strain $\lambda=1.08$ the transmitted light spectra
of L* and R* polarized light are indistinguishable.

\subsection{Effects of strain on the geometry}
\begin{figure}
\resizebox{0.42\textwidth}{!}{\includegraphics{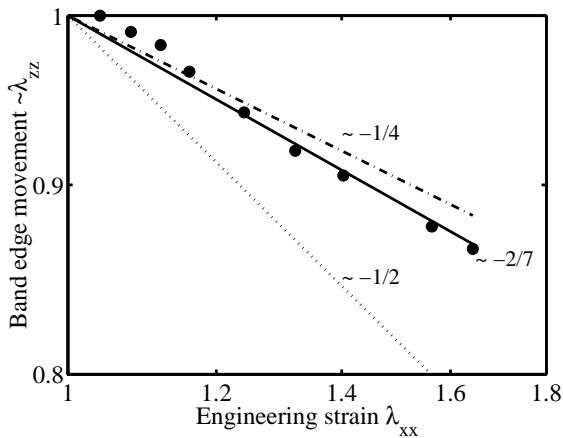}}
\caption{Shift of the reflection gap edge for R* polarized light,
as a function of applied strain $\lambda=\lambda_{xx}$. This is a
measure of the induced contraction $\lambda_{zz}$. Dotted line is
the contraction for an isotropic rubber $\lambda^{-1/2}$, dashed
line is  the theoretical prediction of \cite{WTMM00}
$\lambda^{-1/4}$, solid line is the refined approximation
presented in \cite{BW01}  $\lambda^{-2/7}$. }  \label{strain}
\end{figure}

In an incompressible elastomer, a strain $\lambda_{xx}$ induces a
contraction in $y$ and $z$ directions. For an isotropic rubber one
would expect $\lambda_{yy}=\lambda_{zz}=1/\sqrt{\lambda}$. The
pitch of the cholesteric helix, which is probed by absorption of
R* light, is affinely deformed by the contraction in $z$, and the
shift of the reflection gap edge is a direct measure of this
$z$-contraction. We measured the strain--induced  shift of the
wavelength at which transmission of R* polarized light is
inhibited, Fig.~\ref{figureabsb}(a), and present this data as a
function of the strain $\lambda_{xx}$ in Fig.~\ref{strain}.
It is clear that the data are incompatible with the deformation of
an isotropic rubber. Instead, the data agrees qualitatively with
the prediction from the theory of cholesteric rubber deformation
of \cite{WTMM00} and furthermore we are able to confirm the
validity of the refined approximation reported by \cite{BW01},
which matches the scaling of our data to within our experimental
error. The underlying asymmetry of mechanical properties,
introduced by the cholesteric helix texture, leads to an imbalance
between $y$- and $z$-directions, so effectively the rubber strip
is much stiffer along the pitch axis and the stretched cholesteric
rubber strip contracts much more in the plane of the director and
much less along its thickness: $\lambda_{yy}= \lambda^{-3/4}$ and
$\lambda_{zz}= \lambda^{-1/4}$. A refinement \cite{BW01} of this
theoretical prediction, $\lambda_{yy}\approx \lambda^{-2/7}\equiv
\lambda^{-0.28}$ fits the experimental data in Fig.~\ref{strain}
even better.

\section{Conclusions}
Qualitative agreement is obvious between the experimental and the
theoretical azimuthal intensity profiles, indicating that the
predicted helix coarsening is very likely to be a reasonable
description of the structure of stretched CLCE systems.
Eq.~\ref{phi} has been successfully used to evaluate the patterns
of X-ray scattering at different degrees of imposed extension and
the associated mechanical change $\lambda_{zz}$ reflected in the
gap edge movement.

On increasing strain the cholesteric texture evolves in a
predictable way, coarsening into a sequence of nearly-uniform
director regions, separated by increasingly sharp twist walls.
Deviation from the perfectly sinusoidal modulation of optical
indicatrice in an undistorted cholesteric can be described as a
mixture of higher harmonics and, accordingly, new bandgaps emerge
in both the ``correct'' (R* in our case) and the ``opposite'' (L*)
circular polarisations of incident light. The detailed calculation
in \cite{BW01} gives the particular predictions of photonic
structure and its evolution. Our spectral measurements,
Figs.~\ref{figureabsb}(a,b) can be directly compared to the
bandgaps that occur at the first Brillouin zone boundary. The
general behavior, that is predicted in \cite{BW01} for smaller
wavevectors, would lie in the UV-range for our particular
material, although the fine structure in the L* helicity spectra
may be a signal of that complexity.

Considering first the circular polarization R*, the same helicity
as the cholesteric itself, our main result is the experimental
measure of the scaling of the pitch wavelength as a function of
strain Fig.~\ref{strain}: the observed movement of the band edge
is $\Lambda_{gap}\sim \lambda^{-2/7}$. In addition, an important
observation can be made of the band edge becoming much steeper on
deformation. One could imagine that a number of intrinsic
imperfections, in particular, the possible wandering of the
helical axis away from the sample plane normal, would be reduced
by mechanical strain: the cholesteric texture coarsens, but also
becomes more perfect, defect-free. This might have an important in
lasing and other photonic applications where the increasing of
optical purity of the system is of importance. In the case of
opposite circular polarization, L*, we confirm the absence of a
bandgap for the unstrained material. On deformation we observe the
gradual development of such a gap with no apparent threshold.

Several more detailed issues remain outstanding. In particular,
this relates to the fine structure of transmission spectra for
both R* and L* polarisations (we assume it is due to the selective
reflection, rather than absorption). The fine structure below
450nm is evident, as well as the apparent double-peak modulation,
both seemingly reproducible between different samples and
different stretching events. This modulation, as is the fine
structure on top of it, are not necessarily the noise: one can
clearly resolve the spectra down to $10^{-4}$ level in the
unstrained sample. Intriguing possible explanations include the
effects of randomly quenched disorder (an inherent feature of
liquid crystalline elastomers), further modified by deformation
and affecting the average light propagation through the
cholesteric texture.

To summarise -- we have described the first combined measurements
of optical, structural and mechanical properties of a cholesteric
liquid crystal elastomer and compared with the theoretical
predictions of the deformation and the resulting optical effects.
The behavior of this class of materials under biaxial strain had
been shown to be much simpler and has already found application in
novel photonic devices. The present study confirms our
understanding of the effects of a mechanical field and
demonstrates the surprisingly rich behavior of these materials as
a function of uniaxial strain.

\begin{acknowledgments}
We thank P.A. Bermel and M. Warner for a number of useful
discussions. The preparation of materials has been possible due to
the advice and help of S.T. Kim and H. Finkelmann. This research
has been supported by EPSRC.
\end{acknowledgments}


\end{document}